\documentclass[10pt,conference]{IEEEtran}
\usepackage{cite}
\usepackage{amsmath,amssymb,amsfonts}
\usepackage{algorithmic}
\usepackage{graphicx}
\usepackage{textcomp}
\usepackage{xcolor}
\usepackage[hyphens]{url}
\usepackage{fancyhdr}
\usepackage{hyperref}
\usepackage{subfig}
\usepackage{booktabs}
\usepackage{subfig}
\usepackage{xcolor}

\title{Full System Architecture Modeling for Wearable Egocentric Contextual AI}

\def\hpcacameraready{} 

\newcommand\hpcaauthors{Vincent T. Lee\textsuperscript{$\dagger$}, 
Tanfer Alan\textsuperscript{$\dagger$}, 
Sung Kim\textsuperscript{$\dagger$}, 
Ecenur Ustun\textsuperscript{$\dagger$}, 
Amr Suleiman\textsuperscript{$\dagger$}, 
Ajit Krisshna\textsuperscript{$\ddagger$},\\ 
Tim Balbekov\textsuperscript{$\ddagger$}, 
Armin Alaghi\textsuperscript{$\dagger$}, 
Richard Newcombe\textsuperscript{$\dagger$}}
\newcommand\hpcaaffiliation{Meta Reality Labs Research\textsuperscript{$\dagger$} \\ Meta Reality Labs Silicon\textsuperscript{$\ddagger$}}
\newcommand\hpcaemail{\{vtlee, tanfer, sungmk, ecenurustun, amrsuleiman, ajitkrisshna, timbalb, alaghi, newcombe\}@meta.com}

\author{
  \ifdefined\hpcacameraready
    \IEEEauthorblockN{\hpcaauthors{}}
      \IEEEauthorblockA{
        \hpcaaffiliation{} \\
        \hpcaemail{}
      }
  \else
    \IEEEauthorblockN{\normalsize{HPCA \hpcayear{} Submission
      \textbf{\#\hpcasubmissionnumber{}}} \\
      \IEEEauthorblockA{
        Confidential Draft \\
        Do NOT Distribute!!
      }
    }
  \fi 
}

\fancypagestyle{camerareadyfirstpage}{%
  \fancyhead{}
  
  \fancyhead[C]{
    \ifdefined\aeopen
    \parbox[][12mm][t]{13.5cm}{ }    
    \else
      \ifdefined\aereviewed
      \parbox[][12mm][t]{13.5cm}{ }
      \else
      \ifdefined\aereproduced
      \parbox[][12mm][t]{13.5cm}{ }
      \else
      \parbox[][0mm][t]{13.5cm}{ }
    \fi 
    \fi 
    \fi 
  }
  \fancyfoot[C]{}
}
\begin{document}
\maketitle

\ifdefined\hpcacameraready 
  \thispagestyle{camerareadyfirstpage}
  \pagestyle{empty}
\else
  \thispagestyle{plain}
  \pagestyle{plain}
\fi

\newcommand{\hpcaheight}{0mm}
\ifdefined\eaopen
\renewcommand{\hpcaheight}{12mm}
\fi

\newcommand{\tocite}[1]{\textcolor{brown}{[cite(#1)]}}

\newcommand{\note}[1]{\textcolor{gray}{[#1]}}
\newcommand{\toolname}{PnPSim}

\begin{abstract}
The next generation of human-oriented computing will require always-on, spatially-aware wearable devices to capture egocentric vision and functional primitives (e.g., Where am I? What am I looking at?, etc.). 
These devices will sense an egocentric view of the world around us to observe all human-relevant signals across space and time to construct and maintain a user's personal context. 
This personal context, combined with advanced generative AI, will unlock a powerful new generation of contextual AI personal assistants and applications.
However, designing a wearable system to support contextual AI is a daunting task because of the system's complexity and stringent power constraints due to weight and battery restrictions. 
To understand how to guide design for such systems, this work provides the first complete system architecture view of one such wearable contextual AI system (Aria2), along with the lessons we have learned through the system modeling and design space exploration process.
We show that an \textit{end-to-end full system model view} of such systems is vitally important, as no single component or category overwhelmingly dominates system power. 
This means long-range design decisions and power optimizations need to be made in the full system context to avoid running into limits caused by other system bottlenecks (i.e., Amdahl's law as applied to power) or as bottlenecks change.
Finally, we reflect on lessons and insights for the road ahead, which will be important toward eventually enabling all-day, wearable, contextual AI systems.

\end{abstract}

\maketitle

\section{Introduction}
\label{sec:introduction}

Wearable egocentric-perception devices, such as smart glasses, promise to enable the next generation of human-oriented computing by observing the world from humans' point of view and combining these observations with AI to unlock powerful new personalized contextual artificial intelligence (AI) assistants and capabilities~\cite{abrash_iedm_2021}. 
Similar to the Xerox Alto introduced over 50 years ago, these systems are positioned to revolutionize how we interact with computing devices~\cite{xerox-alto}. 
However, unlike previous generations of human-oriented computing, e.g., personal computers and smartphones, wearable glasses will also have access to egocentric signals such as eye gaze, head pose, and hand positions, which, along with other personal signals, form the user's \textit{personal context} over time. 
This personal context~\cite{google-io-2025} contains significantly richer, more structured, and longer-term personal information which, when combined with generative AI, will unlock more powerful personalized contextual AI applications.

In recent years, generative AI has made significant strides in creating sophisticated models capable of understanding and generating human-like text and images. However, these AI systems are trained on vast datasets that do not include the nuanced personal data that an egocentric wearable device can capture. As a result, this restricts their ability to provide personalized and context-aware assistance. This gap can be bridged by integrating personal context and egocentric signals, i.e., how the user experiences the world, into AI systems.
For instance, AI could observe personal food intake or daily exercise activity, and propose restaurants or routines to meet personal dietary or health goals. This personally tailored AI service to augment and complement human capabilities is what we refer to as contextual AI.

Wearable, always-on devices, such as smart glasses~\cite{rbm, aria2-hardware-specs}, are uniquely positioned to fill this gap by continuously capturing egocentric signals like eye gaze, head pose, and hand positions, along with other personal context signals. The egocentric signals required to construct personal context are computed by a set of \textit{egocentric primitives} (Where am I? What do I see?, etc.). Each primitive has an implementation that takes information sensed from the world around us, computes over it, and generates an egocentric signal. For instance, an egocentric primitive \textit{implementation} such as hand tracking uses outward-facing cameras to sense visual data, compute over it, and generate hand position signals.

From these egocentric signals and workloads, we can derive the device architecture requirements for always-on personal contextual AI. The required input sensing modalities for each primitive define the type, number, and placement of sensors for a wearable device (e.g., RGB camera, inertial measurement units, etc.). The specific algorithm implementations define the architectural resources required to support each workload, as well as the design optimization trade-offs between compute, memory, communication, and other resources. Finally, all of these architectural functional requirements must be unified, operate simultaneously, and  optimized to while satisfying stringent form-factor and battery life restrictions.

\begin{figure*}[h]
\centering
\includegraphics[width=\linewidth]{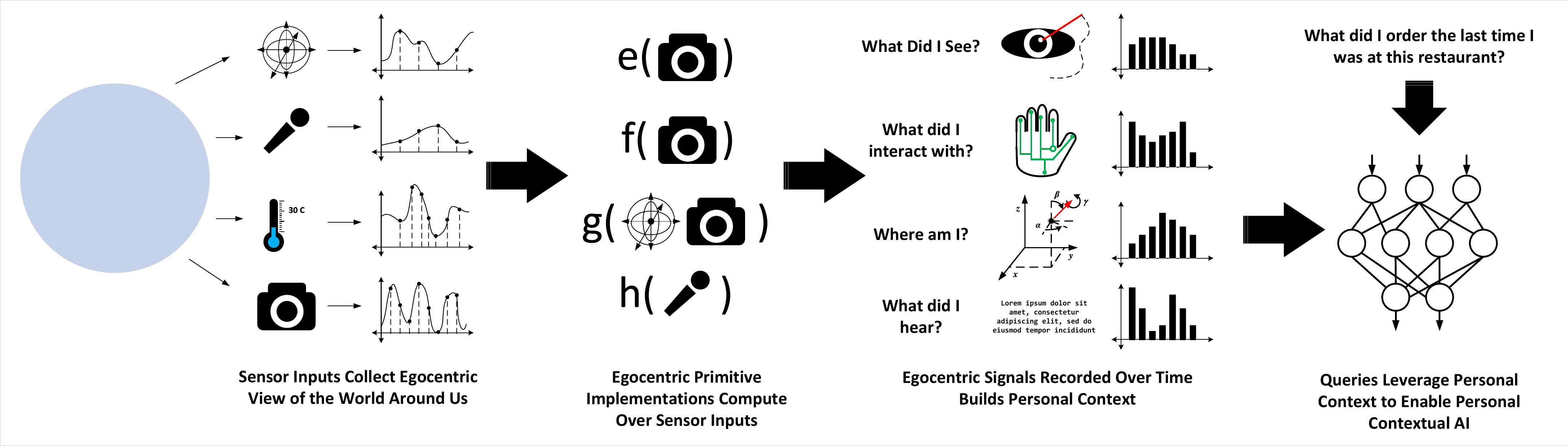}
\caption{Personal context is built by first sensing the world around us, computing over sensor data to extract egocentric signals. Personal context is combined with an AI agent to drive personalized contextual AI services.}
\label{fig:personal_context}
\end{figure*}

More specifically, a comfortable pair of consumer smart glasses today weighs around 50 grams~\cite{rbm, weight-limit}, which restricts the battery size to roughly 3 Whr depending on form-factor limits, functionality, and other device design considerations. The battery will have to last 15 hours to enable all-day operation, which restricts the average power consumption of the \emph{entire} system to about 200 mW. This power target imposes a challenging operating constraint given the variety and sheer number of discrete system components and IPs (which can surpass hundreds) that consume power on a practical wearable system. For instance, in addition to compute, memory, and communication, practical systems must also support various sensors, power management units, output devices (e.g., speakers), and RF connectivity (e.g., wireless radio). Even among compute units, a wearable system may have a variety of SoCs, co-processors, digital signal processors (DSPs), microcontrollers (MCUs), and custom IPs. When optimizing such a system, the operating power cost of all other components adds up, so design decisions and optimizations need to be made in the full system context to properly assess the full system impact of optimizations.

In this work, we provide a view into how we guide design and manage the complex architectural design space to optimize power for a wearable contextual AI device. In particular, we build the first complete system architecture power and performance model for a wearable contextual AI system based on Aria2~\cite{aria2, aria2-init, project-aria-page} and illustrate the important architecture design trade-off behaviors for power optimization. This full system power model also allows us to evaluate the impact of design decisions in full system context and provide insight into the high level architectural trends associated with designing these systems.
We then conduct several design space explorations to illustrate how we evaluate and guide the design decisions, and the some of the key long range challenges. 

We find a key design trade-off is that contextual AI systems must carefully balance the placement of compute for egocentric primitives on-device or off-device to save expensive wireless transmission costs. This is because on-device compute allows us to compress sensor data into smaller egocentric signal representations, effectively trading on-device compute power for reduced wireless offload data movement.
We also find that no single component or category overwhelmingly dominates the system power consumption. This means that to achieve all-day wearable operation, we need to make holistic system design optimizations to reduce system-wide power, instead of applying local (component-level) optimizations that may have limited impact due to Amdahl's limitations applied to power. Finally, we outline some of the key open technical challenges and research problems we anticipate will be important towards building future wearable systems.

The rest of the paper is organized as follows. \autoref{sec:egocentric_primitives} provides an overview of personal context and egocentric primitives. \autoref{sec:system_design} covers the practical architectural requirements. \autoref{sec:modeling} describes our full system modeling methodology. \autoref{sec:result} and \autoref{sec:analysis} show how we evaluate design decisions and trends in the full system context. \autoref{sec:future_work} highlights important research areas for the road ahead.

\section{Contextual AI, Personal Context, and Egocentric Primitives}

\label{sec:egocentric_primitives}

This section provides an overview of personal contextual AI, egocentric primitives, and the architectural resource requirements for a wearable contextual AI system.

\subsection{Personal Context and Contextual AI} 

Despite the impressive capabilities of generative AI, its potential for personal assistance is limited by the lack of direct access to an individual's personal experiences and context. This limitation can result in AI responses to personal queries that are generic, incomplete, or even irrelevant to the user. Furthermore, current generative AI models typically rely on attention mechanisms that focus on short-term patterns and relationships within the data, making it challenging to capture long-term dependencies and personal relationships. To address this, we use \textit{personal context} to provide a structured, compressed representation for organizing user experiences over extended periods of time.

In this work, we define personal context as the collection of all user-relevant experiences across space and time (\autoref{fig:personal_context}). One way to think about personal context is to imagine a narration of a human's life. If we were to describe and narrate what a person did throughout the day, the narration would include the places the person visited, the things they saw and interacted with, the things they said or heard, and so on. As a result, this narration would encode all of the user's relevant personal experiences that would be valuable for answering more personalized queries. By integrating personal context into AI systems, we can enable more effective modeling of long-term dependencies, contextual relationships, and personalized patterns, ultimately leading to more accurate, relevant, and personalized assistance.

\begin{table*}[!htb]
\centering
\caption{A non-exhaustive list of egocentric primitives. Egocentric primitives sense the world around them via various sensing modalities, compute over them using an algorithm implementation, and generate an egocentric signal as output.} 
\begin{tabular}{@{}cccc@{}}
\toprule
Egocentric Primitive & Sensing Modality (Input) & Implementation (Compute) & Egocentric Signal (Output) \\ \midrule
Where am I? & Cameras, GPS, IMU, etc. & Location (Geolocation, Odometry, etc.) & Head Pose / 6 DoF \\
What do I see? & Cameras & Eye Tracking, Image Segmentation & Eye Gaze, Object Label \\
What do I say/hear? & Microphones & Audio Compression, Speech-to-Text & Audio Transcription \\ 
What am I interacting with? & Cameras & Hand Tracking, Object Recognition & Hand Pose, Object Label \\

\bottomrule
\end{tabular}
\label{tab:egocentric-primitives}
\end{table*}

This personal context is especially powerful when combined with an AI agent to enable personalized queries which enables personal contextual AI~\cite{google-io-2025}. 
Personal contextual AI is more powerful because it leverages the necessary context, structure, and information to enable the agent to account for personal experiences in response to queries. 
For instance, instead of asking "where are the keys" (when a user may have observed many keys), a user can ask a more personalized query, "where are \textit{my} keys," which leverages personal context to disambiguate from other potentially incorrect answers.

In theory, all of the user's relevant personal context and information exists in the sensor data which can be observed by an egocentric device throughout their lifetime.
This means that we can potentially feed the entire sensor data stream into an AI agent to drive personal contextual AI.
However, due to system power, storage, and compute limitations, it is impractical to sense and store \emph{all} sensor data across space-time to encode and represent a user's experience as we would quickly run out of storage resources.
Instead, sensor streams must be sufficiently sparsified and compressed to a practically-sized, minimal yet sufficient representation so that they can be offloaded and processed by a backend datacenter.
One such compressed representation can be obtained via egocentric primitives.

\subsection{Egocentric Primitives}

We define a set of egocentric primitives\footnote{There are potentially many variations and may change over time.} that form the basis of personally relevant signals; these signals are collected over space and time to construct and maintain a user's personal context for contextual AI as are defined in (\autoref{tab:egocentric-primitives}). Each egocentric primitive provides the answer to a basic egocentric query about the user, such as "where am I?" or "what do I see?". Ideally, the set of egocentric primitives is complete in that they can be composed and combined to answer any query a user may have; for instance, a combined location and visual query may manifest as "\textit{what} did I see while I was \textit{there}?". An egocentric primitive takes as input a set of sensing modalities such as visual, audio, inertial, etc., computes over them using an algorithm implementation, and generates an egocentric signal as output. Each egocentric primitive is implemented using one or more classes of machine perception and AI algorithms. Each egocentric signal is continuously sensed throughout a user's life to construct a user's personal context across space and over time.

To derive a reasonable set of egocentric primitives, we rely on insights from social science research \cite{cury2019}. This work posits that personal signals are typically what users find most meaningful and useful in their daily lives, such as what they see, hear, or interact with, i.e., the world as experienced from their point of view. We also often want to track and interpret patterns in our daily lives in order to achieve directed goals or outcomes. Egocentric signals are powerful indicators of such user intents, so it is important to capture and store the signals continuously over potentially long periods of time. Finally, many of us find it meaningful to relive or share experiences and build stronger social connections with people we care about, which can also be constructed from egocentric context and signals. From these observations, we define the following set of egocentric primitives to generate the set of personal signals that together constitute a user's personal context.

\noindent \textbf{Where am I?} Spatial reasoning is a key egocentric signal that allows the system and AI agent to reason about a user's location and surrounding environment. 
This includes not only precise localization and head tracking (i.e., millimeter to centimeter-level positional error), but also mapping and indexing of the user's environment. 
Positioning systems like the Global Navigation Satellite System (GNSS) are typically insufficient in terms of positional accuracy and, more importantly, cannot provide egocentric 3D environmental context. 
Instead, to achieve the required accuracy and capture necessary environmental context, this primitive relies on more precise algorithms like simultaneous localization and mapping (SLAM) algorithms~\cite{engel_eccv_2014, murartal_tro_2016}, which leverage sensor fusion across visual and inertial sensing modalities (and can be supplemented with GNSS).

Visual-inertial SLAM is a popular variant of SLAM that takes as input sensor data from one or more outward-facing cameras and inertial measurement units (IMUs). 
The front-end implementation of the algorithm usually consists of a tracking algorithm that uses a combination of visual and inertial information (e.g., VIO~\cite{zheng_icra_2017}, TLIO~\cite{liu_ral_2020}) to generate as output a trajectory encoding translational and rotational coordinates in a local reference frame. 
The back-end mapping system absorbs information (e.g., trajectory and points of interest) from the front-end and creates a map that can be referenced. 
The combination of localization and mapping provides AI agents with sufficiently rich context to reason about queries involving a user's past and present environment. 
Additional sensing modalities such as GNSS, WiFi, and magnetometry can improve localization accuracy, but for this work, we focus on the visual and inertial sensors required for VIO-based SLAM implementations.

\noindent \textbf{What do I see?} Understanding what the user is looking at is an important egocentric primitive because it encodes where the user's attention is. 
While outward-facing cameras are sufficient to capture the general scene around a glasses device, understanding nuanced user attention requires precise information about the user's gaze~\cite{rayner_qjep_2009, spering_arvs_2022}. 
This primitive is typically implemented by eye tracking algorithms~\cite{hansen_tpami_2010}, which determine the gaze of the user and provide key indicators about what they find salient in the environment. 
Gaze is challenging to capture due to several factors, notably variable eye morphology, occlusions (e.g., eyelids and eyelashes), and the high sampling rates required to capture rapid eye movements (e.g., saccades). 
In a glasses form factor, determining gaze requires additional miniaturized inward-facing cameras to capture images used by eye tracking algorithms.

Video-oculography (VOG) methods~\cite{duchowski_ch5_springer_2007} represent the state-of-the-art for eye tracking with high spatial accuracy and precision. 
VOG necessitates inward-facing cameras that capture video of the eyes; some variations will use active infrared illumination to measure and improve robustness under variable lighting conditions. 
Visual landmarks, such as the pupil and reflections on the cornea, are used to build a multidimensional model of the eye and subsequently estimate the point-of-gaze (i.e., gaze vectors). 
Similar to the user's location context, the output signals from eye tracking algorithms are gaze vectors, which provide significant data compression relative to the image data used to calculate them.

\noindent \textbf{What do I hear?} Understanding what the user hears is an important egocentric primitive which captures context about speech, conversations, and environmental sounds. Speech carries context about everyday activities and interpersonal interactions, in addition to serving as a low-friction user interface for glasses. To capture general context from speech and environmental sounds, the device requires one or more air and/or contact microphones to support always-on audio recording. To achieve sufficient directional coverage and fidelity, these microphones need to be physically distributed across different locations on the device. Audio data is then processed by several classes of algorithms, including voice activity detection (VAD) and automatic speech recognition (ASR).

Voice activity detection (VAD)~\cite{rbm-vad} is a lightweight method used to detect whether an audio recording contains speech; it gates computationally intensive downstream speech processing to optimize resource usage. If speech is detected, understanding what the user said or heard is implemented using automatic speech recognition (ASR) algorithms to perform speech-to-text (STT). ASR is typically handled by transformer-based deep learning methods~\cite{shi_icassp_2021, gulati_interspeech_2020}, which can be computationally intensive. The output of ASR is a sequence of tokens or a transcript of what the user said or heard, which can then be used by downstream applications such as keyword detection for voice commands. 

\noindent \textbf{What am I interacting with?} Understanding what a user is interacting with (usually with their hands) provides a compressed encoding of the relevant objects in the environment. This is because objects that the user interacts with are more salient to the user than those they do not interact with (though observed objects are also valuable to capture). The algorithmic implementation of what the user interacts with requires the combination of several algorithms: hand tracking, object detection, and object recognition. Hand tracking algorithms~\cite{ht-dpe} require outward-facing cameras to detect, track, and estimate the pose of each hand. This information can be combined with object detection, object recognition, and gesture recognition algorithms to encode information about what the user interacted with and how.

Object detection and recognition require visual information from high-resolution outward-facing cameras (i.e., RGB camera(s)) to sufficiently capture details about the object. Modern object recognition algorithms are typically implemented using large machine learning models such as DETIC~\cite{detic} and Segment Anything~\cite{sam}, which take visual information as input and output the object label or segmentation. Combined together, these algorithms generate hand pose signals, object labels, and object segmentation as output for a user's personal context.

\section{System Architecture Design Requirements}
\label{sec:system_design}

We now outline the system architecture considerations and requirements for wearable egocentric contextual AI devices.

\begin{figure*}[!h]
    \centering
    \subfloat[Device illustration of sensors suite.]{%
        {        \includegraphics[width=0.47\linewidth]{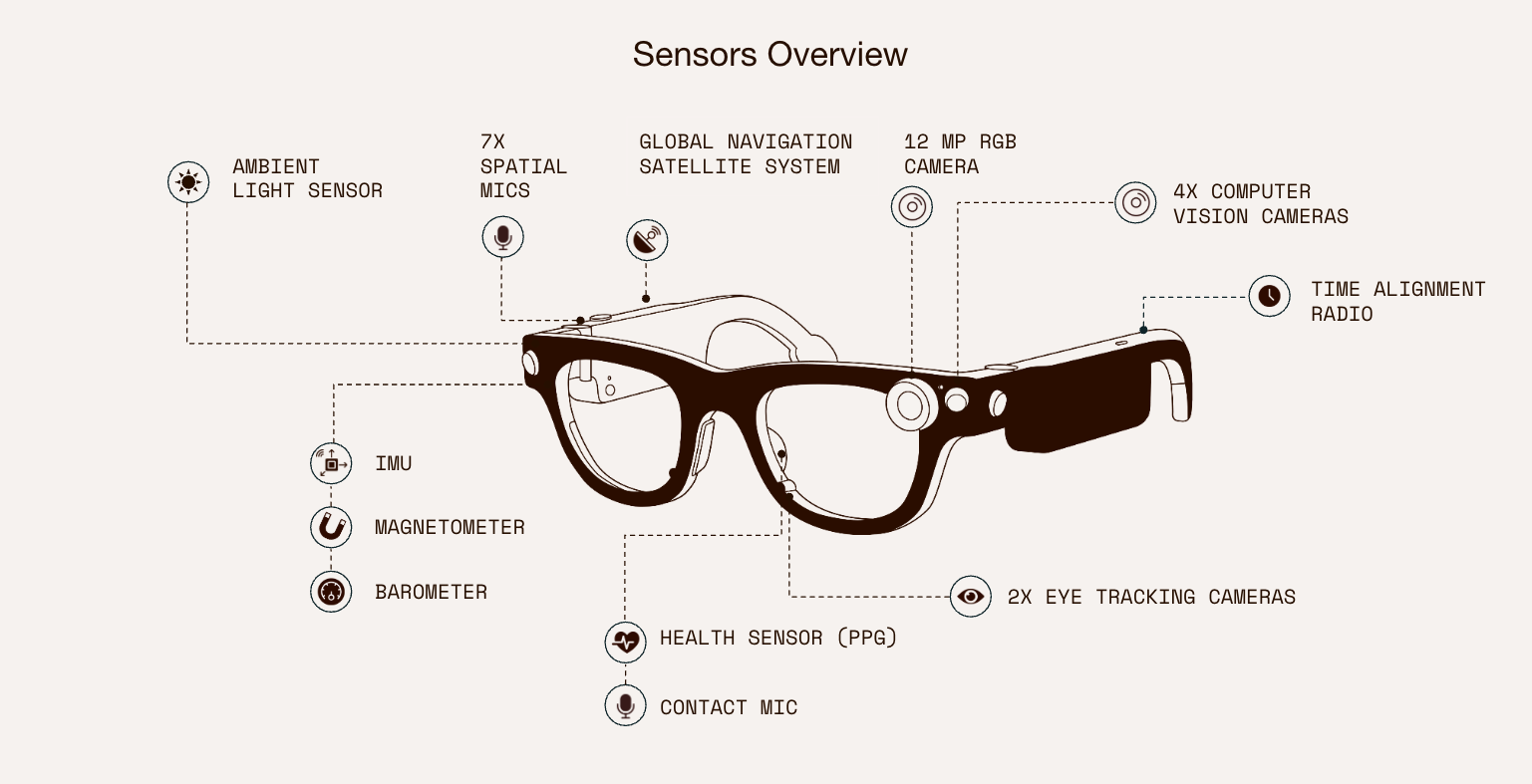}}        
    }
    \hfill
    \subfloat[Abstract system architecture block diagram.]{%
        {        \includegraphics[width=0.47\linewidth]{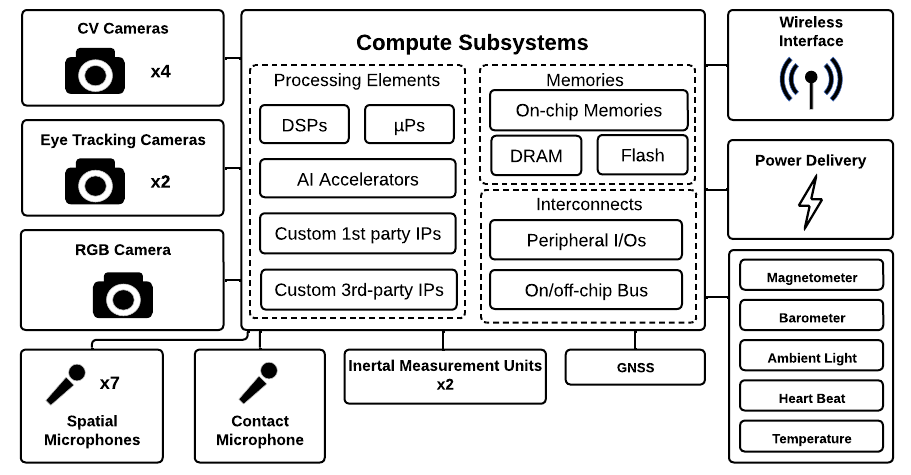}}        
    }
    \caption{Aria2 high level architecture diagram~\cite{aria2-technical-ref}. Compared to existing wearable systems and mobile phones, contextual AI devices like Aria2 have significantly more sensors which better capture an egocentric view of the world. These sensor signals need to be processed efficiently on device prior to offload under a battery constraint much smaller than existing mobile phones.}
    \label{fig:architecture_diagram}
\end{figure*}

\subsection{Architectural Resources}

\noindent \textbf{Sensors.} To build personal context, a contextual AI system needs to observe and understand the world around the user. The system (at a minimum) must support visual, inertial, and acoustic sensors (\autoref{tab:egocentric-primitives}). For visual sensors, the system will need one or more outward-facing color (RGB) and/or greyscale cameras. Systems like Project Aria~\cite{aria} and Aria2~\cite{aria2} have one higher-resolution (and higher-power) RGB point-of-view (POV) camera and several lower-resolution greyscale cameras (also lower power). The RGB cameras support egocentric primitives such as "what do I see?" and "what am I interacting with?". In contrast, smaller, lower-power greyscale cameras distributed around the device provide an enhanced field of view for location estimation algorithms to answer "where am I?". Inertial and satellite data can also be combined with visual input to answer "where am I?", so systems may use one or more inertial measurement units (IMUs) and a GNSS unit. Finally, to support primitives like "what did I hear?", the system will need one or more microphone arrays.

\noindent \textbf{Compute.} Processing sensor data to calculate egocentric signals (or offload from the device) requires a variety of general-purpose and specialized compute devices. The exact composition of these devices depends on which egocentric primitives run on-device and the system architecture. However in all cases, compute capabilities are necessary to process incoming sensor data and push data through the system. Similar to mobile phone architectures, wearable systems can also include specialized processing units like digital signal processors (DSPs) and custom IPs (HWAs) to run tasks as power efficiently as possible. The system may also contain various general-purpose compute cores (e.g., SoCs) and microcontrollers (MCUs) to orchestrate data movement to and from various memory units, as well as to operate drivers and communication stacks.

\noindent \textbf{Memory.} A wearable system will have one or more memory storage units to support data aggregation and stack/swap space for compute. To minimize power, systems must limit the total amount of onboard memory, as each additional bit or storage incurs some idle power overhead. The amount will also need to be sufficient to \textit{concurrently} fulfill the memory requirements of all on-device egocentric primitive implementations which makes the optimization challenging. These aggressive memory size targets limit the deployment of large machine learning models that exceed reasonable embedded memory footprints (i.e., 100s MBs). For example, full-feature, large-scale LLMs (which require 100s of gigabytes) likely cannot be deployed on wearable devices due to space and power limitations. Finally, wearable systems may also support GBs of slower mass storage to aggregate larger amounts of data.

\noindent \textbf{Communication.} All egocentric signals and data that compose personal context will eventually need to be moved off the device to a backend service where it can be stored and queried by contextual AI services. 
This is because over time the size of a users personal context will quickly exceed practical on-device memory storage capacities. 
The system will therefore require both on-chip and off-chip communication devices to move data within a component (e.g., a coprocessor) and between components (e.g., sensors to compute). 
Additionally, the system will need to support radio communication components such as Bluetooth and/or WiFi to transmit egocentric signals off the device. 
These radio components will also require auxiliary supporting services, such as link maintenance and scanning, to maintain connectivity, which consumes power.

\noindent \textbf{Power Delivery.}
Voltage regulators and power management integrated circuits (PMICs) play a important role in mobile devices, including wearable glasses. 
Different components in the system (e.g., at the PCB and silicon power domain levels) require a variety of supply voltages with different load current requirements, necessitating regulation down (or up) from the battery voltage rail. 
However, load regulation and voltage conversion are inherently inefficient, consuming power in addition to that delivered to the load; this loss is characterized by an \textit{efficiency factor}. 
For example, an efficiency factor of 75\% implies that the PMIC or regulator draws 100 mW of power to deliver 75 mW of "useful" power to the load. 
As a result, every component in the system effectively incurs additional power and energy overhead due to power delivery and voltage regulation.

\subsection{Weight, Battery Life, and Power}

Contextual AI glasses must be physically comparable to standard glasses to be comfortable and socially acceptable. 
Glasses typically weigh no more than 15-40 grams~\cite{eyeglass-weight, weight-limit}. 
While some frame material can be substituted for electronics, most of it must be preserved to maintain structural integrity. 
This means any additional electronics will be mostly additive to the weight of the glasses.
Existing smart glass offerings like Ray-Ban Meta (RBM) weigh about 50–60 grams (the majority due to the frame). 
Since plastic frames on average weigh 20-30 grams~\cite{eyeglass-weight}), we can roughly extrapolate a comfortable additive weight of $\approx$20-30 grams for the electronic components.

The battery is typically the heaviest electronic component in the device and limits the power and operating time. 
The energy density of a typical lithium-ion battery ranges between 200–300 mWh/g~\cite{zu_lithium_2011, sun_iphone_2019, zhao_lithium_2021}.  
If we assume the battery is limited to about half the 20–30g additive weight budget for electronics, the energy density of lithium-ion battery chemistry permits a capacity of roughly 3 Wh (10 g @ 300 mWh/g). 
This is roughly an order of magnitude (4–12$\times$) less capacity compared to a typical smartphone battery (about \textasciitilde 10 Wh~\cite{iphone-battery}).

The battery restriction imposes a key design constraint because it limits the average operating power of the system. 
To build personal context over a user's lifetime, the device must be always on, requiring operation of at least 15 hours on a single charge. 
Assuming a 3 Wh capacity, the all-day requirement translates to a \textasciitilde 200 mW ceiling on average power dissipation for the entire glasses system. 
This means that all egocentric primitives, supporting functions, and overheads must operate on average below this power target which make the design of such a system extremely challenging.

\section{Power Modeling End-to-End Wearable Architectures}

\label{sec:modeling}

\subsection{Simulation Methodology}

Designing a contextual AI system that simultaneously supports and operates all egocentric primitives under strict power and weight limitations is a complex architectural challenge. 
To enable design space exploration over such a large space and manage these complexities, we built a complete system model to unify workload requirements with system architecture resources which consume power. 
To achieve this, we developed an event-driven (using simpy~\cite{simpy_docs}) system architecture simulator \toolname{} in Python to construct a full system power and performance model of a contextual AI system (Aria2~\cite{aria2}). 
We use \toolname{} over other existing simulation tools due to the system's sheer complexity and scale, which requires combining profiling data and results from a wide range of design automation tools, datasheets, measurements, and proprietary data sources.

More specifically, \toolname{} takes a set of workload specifications and a system architecture specification as input and generates a full system power and performance estimate as output. 
To model each application workload, we construct a dataflow dependency graph or taskgraph for each egocentric primitive implementation.
Each task in the dataflow graph tasks specifies the architectural resource requirements such as the compute time or cycles, device type, other task dependencies, data allocation/movement costs, etc. 
The performance characteristics and architectural resource requirements for each task are constructed from various EDA/CAD simulation tools (Synopsys, Cadence, etc.) and profiling results.
The task resource requirements are then abstracted into a standardized \toolname{}-compatible specification which allows us to model contention and also combine IP-specific internal performance data and tools.
These resource requirements are then used to map each task onto the necessary architectural resources to simulate simultaneous execution of every workload.

The system architecture model is specified as a block diagram consisting of a set of devices and their connectivity. 
Each device specifies the architectural resources it provides (e.g., compute, memory, communication): hardware IPs provide compute resources, memory modules provide storage resources, bus interfaces provide communication resources, and so on. 
Each device has a state-based performance model which records the duty cycle in each state. 
For instance, a hardware IP will be in an idle state when not used and active state when it is allocated to service a compute task.
The duty cycles are then combined with state-based, throughput-based (ex., energy per byte), and device-specific power models.
Power modeling data for individual devices and subsystems is derived from pre-silicon power simulation estimates, isolated power measurements on individual IPs/subsystems, vendor data sheets, and IP-specific analytical models.
We have generally found for silicon IPs that the power modeling data tracks within about 10\% of the actual power.

\toolname{} then simulates the workload run time by scheduling each taskgraph's resource requirements against available architecture resources while also accounting for resource contention between workloads.
The simulation then generates an estimated duty cycle resource utilization as well as other system architecture telemetry such as device state traces (e.g., active, idle) and the amount of transmitted data. 
The system telemetry is used to drive power estimation for each device, which are aggregated bottom-up to generate a full system power estimate. 
When reporting numbers from \toolname{}, we round power results to two significant figures before normalizing to percentages; this is to protect confidential information without altering general architectural trends and design challenges. 
These power and performance model estimates can then be compared across different operating points and designs to drive rapid design space explorations to provide insight into trends and inform system design.

\begin{table}[t]
    \centering
    \caption{Sensor model parameters based on~\cite{aria}. Settings will vary across algorithm variants and device architectures.}
    \label{tab:sensor_parameters}
    \begin{tabular}{ccc} \toprule
    Sensor & Resolution & FPS \\ \midrule
    POV RGB Camera & 1440 x 1440 & 5 Hz \\
    Greyscale Cameras & 640 x 480 & 30 Hz \\
    ET Cameras & 320 x 240 & 30 Hz \\
    IMU & 1 x 6 & 800 Hz \\
    Microphones & 1 & 48 kHz \\
    GNSS & 1 & 1 Hz \\
    Magnetometer & 1 & 100 Hz \\ 
    Barometer & 1 & 50 Hz \\
    \bottomrule
    \end{tabular}
\end{table}

\subsection{Architecture Model}

\begin{figure*}[h!]
    \centering
    \begin{subfloat}[Full offload power composition. Sensor data is compressed and offloaded and egocentric signals are computed off device. \label{fig:compressed_sensing_power}]{        \includegraphics[width=0.45\linewidth, trim={0 3cm 0 3cm}]{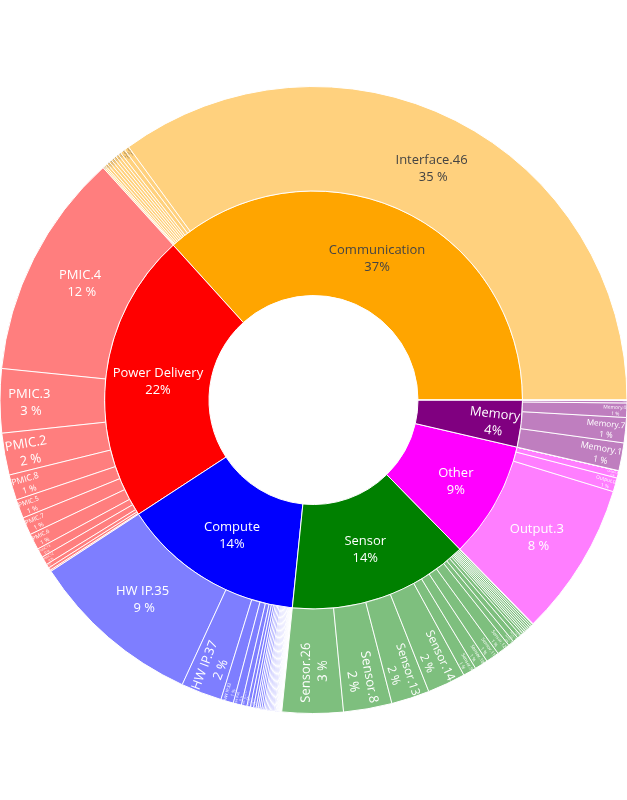}}
    \end{subfloat}%
    \hfill
    \begin{subfloat}[On-device compute power composition. All egocentric signals are computed on-device then transmitted off device.\label{fig:on-device-compute}]{        \includegraphics[width=0.45\linewidth, trim={0 3cm 0 3cm}]{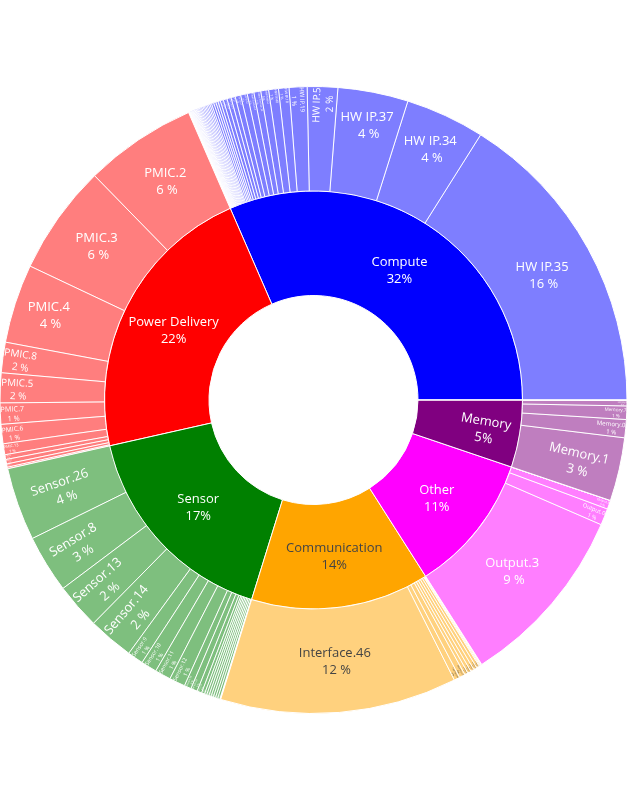}}
    \end{subfloat}

    \caption{
        Power composition by device category (normalized out of 100\%). On-device computation is 16\% lower power relative to the full offload configuration.}
        Power bottlenecks change depending on the use case highlighting the importance of a full system view of power.
        Power composition will vary depending on architecture, egocentric implementation, and system maturity.
        Device names and power values are obfuscated to protect confidential information.
    \label{fig:dsa_comparison_points}
\end{figure*}

A high level architectural system block diagram is shown in ~\autoref{fig:architecture_diagram} based on~\cite{aria2-technical-ref}.
One of the key notable architecture differences for our contextual AI wearable systems compared to mobile devices and even earlier generation wearable systems like RBM is the number of egocentric sensors and modalities.
While mobile phones may have a comparable suite of sensor modalities, they are not purpose-built to capture the world from an egocentric perspective.
Contextual AI devices such as Aria2 includes inertial measurement units, microphones, RGB, inward- and outward-facing cameras, as well as Bluetooth and WiFi transceivers, GNSS, barometer, magnetometer, and others~\cite{aria2-hardware-specs}.
This significantly diversifies the number of devices which need to be accounted for when modeling the full system power.

In terms of compute subsystems, Aria2 has an onboard coprocessor~\cite{meta-xr-custom-silicon} which handles various hardware acceleration support.
The device’s coprocessor supports on-device compression for image/video using H265 HEVC and audio data using OPUS encoders. 
It also features hardware-accelerated support for machine perception, including 3D articulated hand tracking, eye-tracking with gaze per eye output, and advanced signals such as pupil diameter and blink detection, as well as 6DoF localization.
The device also has both on-chip and off-chip memories, as well as a mass storage flash unit similar to RBM and Aria.
To protect confidential information, we do not provide detailed architectural connectivity and subsystem specifications; however, this will not be necessary to illustrate architectural power trends.

\section{System-Level Experiments}
\label{sec:result}

This section explores how different combinations of on-device and offloaded egocentric primitives impact power. 

\subsection{Wearable System Power Composition}
\label{sec:power_composition}

To understand the system complexity and power trade-offs, we first examine the power breakdown of a complete wearable system - Aria2~\cite{aria2} - and how it varies across operating points. We evaluate two design configurations: (1) full streaming all sensor data and computing egocentric signals off-device (to a backend server), and (2) on-device compute of all feasible egocentric signals before being uploaded.

Across all experiments, we assume a 10:1 compression ratio for raw camera sensor data prior to upload to a backend server\footnote{In practice compression rates will vary with settings and data}. 
For RGB, we apply 2x2 binning to frames, reducing the original 2880x2880 resolution to 1440x1440 in our experiments~\cite{aria}. 
For wireless transmissions, we assume a reasonable good link connection (MCS8).

\subsubsection{Full Offload Scenario} 
\label{sec:full-offload}

In this configuration, the device simply captures data from all sensor modalities and transmits it to a backend for computation; we refer to this as the \textit{full offload} scenario.
In this case, no egocentric signals are computed on the device (i.e., all data is fully offloaded), and the device effectively operates as a recording and streaming device. 
More precisely, records and aggregates sensor streams, and then compresses and uploads the data the server via wireless interfaces off the device. The backend infrastructure then processes and computes the egocentric signals to build the personal context required to drive contextual AI.

The average operating power composition for this full offload configuration is shown in Figure  \autoref{fig:compressed_sensing_power}.
The device power breakdown shows that a few components consume significant power, but no single component or category dominates total power; we also note that the power still far exceeds the target 200 mW always on operating target.
Among the larger power consumers is the wireless interface, which use substantial power due to the amount of compressed sensor data that must be periodically uploaded (\autoref{tab:sensor_parameters}). 
Since wireless power roughly depends on the channel bandwidth, this highlights the importance of compressing data on-device as much as possible to reduce wireless link utilization and overall system power.

\subsubsection{Full On-Device Compute} 
\label{sec:on-device-compute}

We also evaluate a configuration where all egocentric primitives are computed on-device (i.e., hand tracking, eye tracking, VIO, and ASR) to illustrate the power impact on overall system power; we refer to this scenario as \textit{full on-device compute}. 
In this architecture, sensor data is aggregated in memory and processed by onboard IPs to generate egocentric context signals for hand tracking, eye tracking, ASR, and VIO. 
The system however still captures and compresses RGB images using standard compression techniques as object recognition models such as DETIC~\cite{detic} are too large to fit on-device. 
Computing egocentric signals on-device effectively applies algorithmic compression to sensor data since the egocentric signals are typically much smaller than the input sensor streams. 
As a result, the necessary data to upload to build personal context is reduced compared to the full offload scenarios above which reduces wireless link utilization and power.

Figure \autoref{fig:on-device-compute} shows the power composition for this full on-device compute scenario; this configuration actually consumes \textasciitilde 20\% more power than the full offload scenario which shows moving egocentric workloads on device is not always power optimal depending on the implementation.
Compared to the full offload scenario, the on-device computation of egocentric signals reduces the share of wireless communication power by trading additional compute power compared to the full offload scenario previously. 
However, even though \textit{most} sensor signals are compressed by several orders of magnitude (into egocentric signals) the total communication power gains are still limited.
This is because of the RGB frame offloads and overheads associated with maintaining the link which limits the power reduction of the wireless communication link. 
The sensors and outputs, on the other hand, have the same fixed power cost, imposing an Amdahl's limitation on the power impact of other trade-offs in the system. 
This illustrates the importance of a full system model to track and understand how bottlenecks and Amdahl's limitations applied to power shift as the system is optimized.

\subsubsection{Amdahl's Limitations for Power} 
\label{sec:amdahls_limits}

Across both scenarios, there is generally no single component nor category of components that consistently dominates system power. 
\autoref{tab:distribution} shows the distribution of power across components in the system for the on-device compute operating conditions from Figure \autoref{fig:on-device-compute}. 
The data shows a heavily-tailed component power distribution, where most individual components consume only small fractions of the total system power. 
For example, in our system model 129 components individually consume less than 1\% of total system power but collectively account for 17.49\% of the overall system power.

Even if we could achieve orders of magnitude improvement for the two highest power-consuming components, the overall system power improvement would be limited to $\sim$$1.6\times$ since the remaining 143 components consume 61.60\% of the power (1/0.616$~\approx$1.6$\times$). 
In other words, the heavily-tailed component power distribution imposes an Amdahl's limitation on power improvements of individual components.
This means to bring down the overall system power meaningfully, designers will have to holistically combine design optimizations from across the stack to reduce power across multiple system components.

\begin{table}[b]
    \centering
    \caption{Component power distribution (cumulative) for on-device compute. Many consume little power but cumulatively add up.}
    \label{tab:distribution}
    \begin{tabular}{c|c|c} \toprule
    Component Power & \# Component & \% Total Power \\ \midrule
    $ \leq 0.1\% $                    & 82 & $1.47\%$ \\ 
    $\leq 0.5\%$    & 118 & $9.47\%$ \\
    $\leq 1\%$      & 129 & $17.49\%$ \\
    $\leq 5\%$        & 140 & $43.29\%$ \\
    $\leq 10\%$       & 143 & $61.60\%$ \\
    $\leq 25\%$      & 145 & $100\%$\\
    \bottomrule
    \end{tabular}
\end{table}

\subsection{Architecture Design Space Exploration}
\begin{figure*}[h]
    \centering
    \includegraphics[trim={0 0 0 0.5cm},clip, width=\linewidth]{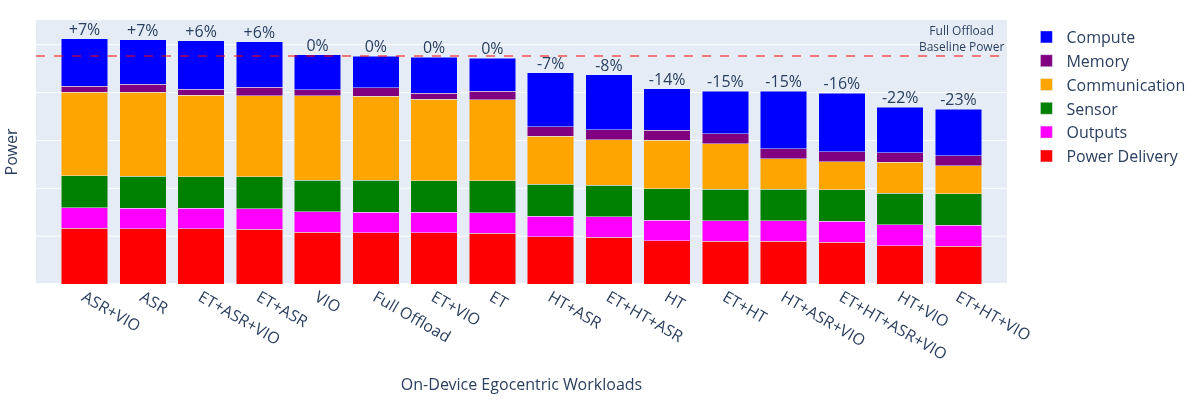}
    \caption{Power composition for different subsets of egocentric signals computed on device. Egocentric signals computed on device upload sensor data and compute signals off device on a backend server. On-device compute trade-offs compute power for reduced communication power. Actual values for will vary with algorithm variant, system architecture, and maturity.
    }
    \label{fig:egocentric-primitives-dsa}
\end{figure*}

One of the key design knobs for contextual AI wearable systems is whether to compute egocentric signals on-device or offload computation to a backend server. 
To evaluate the impact and trade-offs on system power, we conduct a design space exploration over system configurations where different combinations of egocentric primitives are run on- and off-device. 
For egocentric primitives computed off-device, we aggregate the necessary sensor streams, compress, and offload the data. 
Again, we note that RGB frames are always offloaded from the device across all configurations since object detection models like DETIC~\cite{detic} currently cannot be feasibly placed on-device.
\autoref{fig:egocentric-primitives-dsa} shows the relative full system power consumption for each configuration, decomposed by resource category. 

The first key insight is that moving each egocentric primitive on-device does not always reduce power; this highlights the importance of evaluating the trade-off between compute and communication power when determining what to run on-device. 
For instance, hand tracking on device improves power consumption by 14\% since the savings from communication substantially outweighs the additional required compute power.
Eye tracking, on the other hand, roughly breaks even in terms of power by trading off additional on-device compute to reduce communication cost.
ASR actually shows that the overall system power increases by 7\%; this is because compressed audio streams are already relatively low bandwidth compared to image sensors. 
For instance an audio stream consumes around 128 Kbps after compression while a 512 $\times$ 512 image at 30 fps with 8-bit pixels at 10:1 compression requires 6.3 Mbps.
As a result, the power savings opportunity from reducing audio bandwidth communication is small compared to other sensor modalities.
Finally, running only VIO on-device marginally increases system power because outward facing camera frames are shared with hand tracking so the frames still need to be offloaded from the device.

The second key insight is that assessing whether to compute egocentric primitives on-device requires a full system view, as some on-device resources are shared across primitives. 
For instance, the combined savings (-22\% power) of running VIO and hand tracking on-device is better than individually running only hand tracking (-14\% power) or only VIO (+1\% power). 
This is because, in our system architecture, the outward-facing cameras for VIO and hand tracking are shared (albeit at different operating frequencies); to completely eliminate the need to offload outward-facing camera frames, both VIO and hand tracking need to run on-device.
Other shared system resources like on-device memories, on-chip communication, and top-level SoC components—including control circuitry and clocks—also need to be shared across workloads imposing an Amdahl's limitation to power gains. 
These cross-workload interactions of shared components and costs highlight, once again, the importance of assessing system design trade-offs in an end-to-end, full system context to understand how bottlenecks shift around the system.

Finally, we highlight that power-optimal solutions and design trade-offs are likely to change over time as algorithms and device architectures mature. 
Over time, we expect improvements to the underlying architecture to enhance silicon power efficiency due to design improvements and technology node scaling. 
Similarly, we expect the set of egocentric primitives to shift and mature, which in turn may impact the specific sensor specifications and modalities on a wearable system. 
These changes are expected to shift the relative magnitude of the compute-communication trade-off presented in \autoref{fig:egocentric-primitives-dsa}. 
However, we expect high level trends like the compute-communication trade-offs to still hold long term and need to be continuously evaluated within the full system context to assess the power impact.
\section{Current Industrial Trends}

\label{sec:analysis}

Designing wearable systems is a long-term industry effort where we will need to continuously assess how power bottlenecks may shift over time amid technology trends.

\subsection{Technology Scaling}

\label{sec:tech_scaling}

Technology scaling is an industry trend that we expect to continue improving the energy and power efficiency of wearable systems. 
As process technology geometries continue to shrink, system power will generally decrease (at iso-performance) orthogonal to system architecture design decisions. 
However, due to the variety of different devices in our system, the impact of technology scaling will be non-uniform. 
For instance, the static leakage and dynamic power for digital logic will scale at different rates compared to sensors which include both analog and digital silicon. 
To evaluate the impact of technology scaling, we decomposed our power models by semiconductor process and power type. 
Using historical technology scaling factors from major semiconductor manufacturers, we construct a predictive technology process scaling model for each future process node. 
To protect proprietary IP, we obfuscate this data by rounding and then extrapolate with curve fitting to project future scaling factors. 
We estimate the release cadence of future process technology nodes using the historical average intervals between release cycles of major semiconductor manufacturers (approximately every 2 years).

\begin{figure}[b]
\centering
\includegraphics[trim={0 0cm 0 0cm}, clip, width=\linewidth]{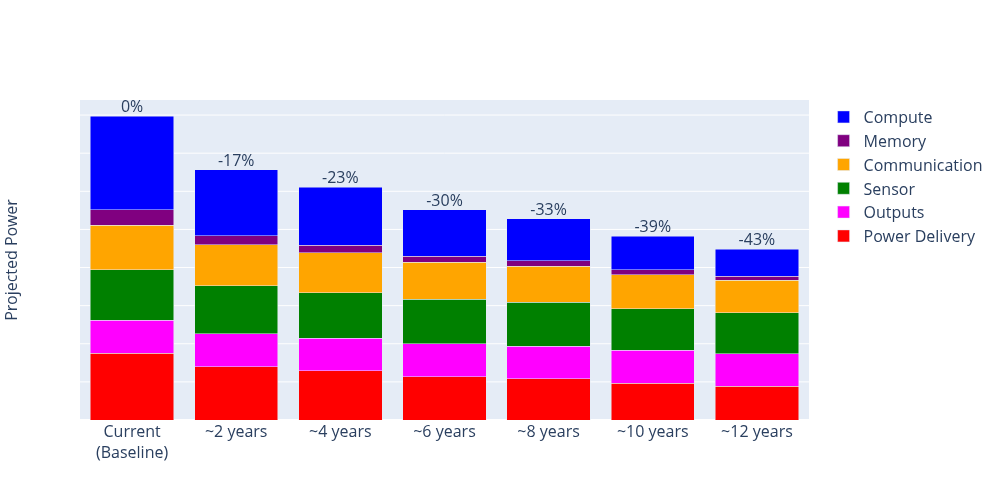}
\caption{Impact of technology scaling for on-device compute case. Digital devices scale better than analog components which make analog device bottlenecks more acute over time.}
\label{fig:tech_scaling}
\end{figure}

\autoref{fig:tech_scaling} shows the system-level power impact of our projected technology scaling on the \textit{full on-device compute} scenario from \autoref{sec:on-device-compute}. 
As expected, technology scaling has an asymmetric effect on system power across different components. 
For example, analog components such as sensor front-ends, PMICs, and LEDs are not projected in our model to scale as well as digital components. 
Moreover, the gains from reductions in dynamic and leakage power in digital logic are limited by components that do not scale as effectively. 
Over time, components that scale less will become increasingly acute bottlenecks in the system, necessitating that future system designs and algorithms take this into account. 
This underscores the importance of evaluating design changes in full system context, as the bottlenecks shift over time.

\subsection{Wireless Compression}

Wireless data offload power constitutes one of the larger shares of system power and, as shown in \autoref{fig:tech_scaling}, is not projected to improve as quickly over time as other system components due to the analog and RF elements. 
This means that, over time, wireless offload power is expected to become a more acute bottleneck as other system components gradually improve with optimization and scaling. 
This further emphasizes the need for co-optimizing wireless communication power with the design and implementation of egocentric primitives as part of the long range research strategy for wearable systems.

The most straightforward, near-term design optimization for reducing wireless power is to compress and/or batch data upload sizes to minimize bandwidth utilization. 
To evaluate the potential power savings behavior, we conduct a sensitivity analysis over different hypothetical compression rate targets to explore the overall system power impact.
In this experiment, we use the streaming \textit{full offload} configuration (\autoref{sec:full-offload}) as the baseline scenario and apply various hypothetical levels of data compression [1:1, 2:1, 4:1, 8:1, 16:1, 32:1, 64:1, and 128:1] and sensor frame rate reduction [1$\times$, 2$\times$, 4$\times$, 8$\times$, 16$\times$, 32$\times$] to explore the impact on wireless link power. 
Such data compression levels can be achieved via computing egocentric signals on-device, spatial sparsification techniques like cropping regions-of-interest (ROI), or frame rate reduction by exploiting temporal sparsity.

\begin{figure}[b]
\centering
\includegraphics[width=\linewidth, trim={0 1cm 0 2cm}, clip]{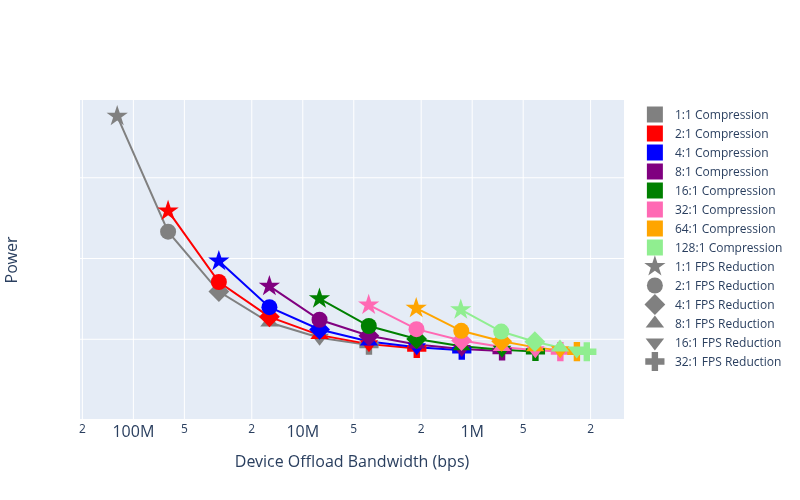}
\caption{System power for streaming compressed sensor data at different compression rates and FPS.}
\label{fig:compression_plot}
\end{figure}

The high level trends of different compression levels is shown in \autoref{fig:compression_plot}. 
The results demonstrate that increasing compression levels asymptotically reduce overall system power due to unavoidable overheads associated with maintaining the link which imposes an Amdahl's limitation for power gain.
At bandwidths below 1 Mbps, it is possible to change to bluetooth paired with a companion device if the wireless link is not reliable which can save power. 
Overall, a key architectural optimization for wearable systems is to compress and reduce data movement to minimize wireless link power

\subsection{Power Delivery}

Power delivery losses can account for a substantial proportion of total system power depending on PMIC efficiencies. 
While absolute power losses improve proportionally to reductions elsewhere in the system, under current trends, we do not expect the \textit{proportion} of losses (characterized by regulator efficiency) to improve significantly for the wearable system studied here. 
This is due to two key reasons: \textbf{1)} form factor and industrial design (ID), and \textbf{2)} battery technology trends.

In terms of form-factor and ID, on-device regulator efficiencies are largely constrained by the size and characteristics of the passive components (e.g., capacitors and inductors) that can be incorporated. 
As system components are miniaturized—e.g., due to improvements in packaging or integration—free physical volume can be recovered to improve ID or utilized to increase regulator efficiency via larger (optimal) passives. 
The form factors of glasses such as Ray-Ban Meta, while acceptable, still have room for improvement in terms of ID. 
Many current wearable systems prioritize user wearability, utilizing savings in physical volume to improve ID while keeping PMIC efficiencies roughly constant.

In terms of battery efficiency, power delivery efficiency is also proportional to the difference between regulator input and output voltages; that is, higher efficiencies can be achieved "for free" if required output voltages are similar to the battery voltage. 
However, supply voltages for custom silicon are trending downward to reduce power, while voltages supplied by batteries are dictated by battery internal chemistry. 
Barring breakthroughs in battery technology that simultaneously enable high energy density, low internal resistance, and low output voltage, we expect the gap between battery supply voltages and downstream voltages to remain relatively constant (greater than 2–3 V gap for digital silicon and greater than 1 V gap for RF/mixed signal). 
Thus, under current trends, we can expect power delivery to continue consuming around the 20\% of total system power.

\section{Looking Forward and the Road Ahead}

\label{sec:future_work}

This work presents the most complete system model to date for wearable contextual AI systems, however it still falls short of the 15 hour always-on operating target.
The road ahead will expect innovations from across the stack to come together to continue to reduce the system power.

\subsection{Full-System Modeling and Design Automation}

\noindent \textbf{Driver Modeling.} Device drivers are needed to orchestrate and operate hardware IPs. 
As a result, they account for an important software overhead which remains challenging to model in the absence of concrete implementations. 
Heterogeneous architectures like the one in this work require many different drivers to work in concert due to the number of hardware IPs, DSPs, and other devices in the system. 
Existing hardware IPs (from similar device) will often have readily available driver performance overhead models, while new IPs typically will not. 
However, modeling the impact of drivers on power and performance in the absence of a concrete implementation remains a challenging task, and design automation solutions still leave much to be desired. 
As a result, driver modeling tools to provide faster and more precise estimation of overheads is a valuable technical gap that system modeling research can fill for wearable systems.

\noindent \textbf{Pre-RTL Power Estimation.} One way to improve the system power is to increase the number of custom silicon hardware IPs to make improvements in on-device compute power efficiency. 
By trading less on-device compute power for reduced communication power, the compute-communication trade-offs can be made more favorable for on-device compute. 
However, custom IP opens up a large design space, which requires time-consuming design space explorations, especially if evaluations need to go to post-place\&route. 
Pre-RTL power estimation modeling enables trading-off modeling accuracy for speed. 
This allows designers to more quickly and broadly assess solutions and prioritize IPs with higher power savings before investing significant engineering effort in IP design. 
Towards this end, combining more accurate, fast, and fully-automated pre-RTL power and performance estimation tools~\cite{aladdin, hls_gnn, hls_intmul} with full system modeling should enable more complete and efficient optimization for complex wearable systems.

\noindent \textbf{Cross-target Power and Performance Estimation.} Wearable systems include a heterogeneous mix of SoC general-purpose cores, digital signal processors (DSPs), domain-specific accelerators (e.g., ML accelerators), hardware IPs, and more. 
During development, applications are typically prototyped on general-purpose cores before being manually optimized for a specific embedded target. 
Compiling an optimizing across different targets remains a challenging task because embedded code usually includes architecture-specific optimizations. 
For example, DSP code needs to be vectorized and will have different power and performance compared to embedded CPU or even other DSP targets due to different vector widths. 
This reduces the agility with which systems designers can reason about the power and performance trade-offs when running an application on different design targets. 
Therefore, enabling automation to more seamlessly explore how workloads perform across heterogeneous computing targets is valuable to enable rapid design space evaluation for questions like, ``What if we ran task X on compute device Y?''

\subsection{Software and Hardware Optimizations}

\noindent \textbf{Temporal and Spatial Sparsity.} Exploiting sparsity is a key technique that can reduce the amount of data that must be compressed, processed, and communicated. 
This makes it a valuable class of optimizations because it reduces power consumption along the entire data path. 
Capturing less data on the sensor reduces the duty cycle and, in turn, decreases the amount of data moving through the system. 
This also reduces compute processing, storage overhead, and off-ship communication, resulting in power savings across the system. 
Sparsity generally manifests as either temporal or spatial. 
Temporal sparsity can be as simple as reducing the frame rate or more complex like using delta compression. 
Spatial sparsity includes techniques like extracting regions of interest from frames or sparse capture modes. 
Sparsification techniques that do not change key algorithmic performance metrics are preferred because they transparently lower overall power. 
However, if sparsification techniques do affect algorithm metrics, they must be accompanied by co-designed algorithm changes, which will likely alter compute resource requirements.

\noindent \textbf{Partial Execution Offload.} Partial execution of an egocentric primitive on the device involves identifying a split point in the application and running all compute before it on-device and offloading the remainder.
Whether this technique will yield power savings will depend heavily on the algorithm characteristics, intermediary representation at the split point, and distribution of compute before and after the split point.
The ideal application for this type of approach would exhibit a highly compressed intermediary representation followed by significant off-device computation.
This would allow the partial offload to save wireless power from compression, as well as save on-device compute power by partially offloading the remaining computation.
To better identify which algorithms exhibit this behavior will require more extensive workload profiling and analysis to assess for these behaviors which we leave as an open question for future research. 

\noindent \textbf{Power Delivery.} Power delivery accounts for a substantial fraction of total system power in due to the inherent costs of voltage conversion and load regulation. 
For wearable systems, power delivery is particularly challenging because form factor constraints limit the size and height of passive components (e.g., filter capacitors and inductors). 
Researchers should focus not only on the design and optimization of DC-DC converters to improve figures of merit (e.g., efficiency, response time, ripple) but also on system-level optimization of the entire power delivery solution. 
While the former is a well-established problem in the circuit design community, research efforts should continue with a focus on area-efficient solutions targeting low-power, low-voltage applications. 
The latter is a less-established multi-objective optimization problem, encompassing relationships between the battery, multiple levels of regulation, leaf-level voltage-current loads, and \textit{physical constraints} such as board area and physical volume.

\noindent \textbf{Novel Full System Optimizations.} As shown in \autoref{sec:amdahls_limits}, the power composition is not overwhelmingly dominated by any single device or device category. 
This means hyper-optimizing individual components or category of devices will have limited impact on overall system power. 
Instead, power optimization efforts need to go beyond individual components and assess impact in full system context across sensors, compute resources, memories, and other workloads. 
In other words, the most impactful long range research ideas and technologies for reducing power will likely be those that lower power across multiple subsystems and/or components. 
Furthermore, optimizing all components in isolation does not yield the optimal solution. Instead, algorithms and components need to be co-optimized with hardware in the loop at design time. For example, determining a partial-offload partitioning of an algorithm cannot be performed in isolation as other algorithms or components may limit the benefits.
Going forward, we expect that such holistic power optimization techniques will be increasingly important and complementary to optimization of near-term big ticket components like wireless transmission.

\subsection{Industrial Design and Thermals}

A glasses-like device must have a socially acceptable form factor while also complying with constraints on weight and thermals to ensure comfort during the desired interaction duration. 
While power is a key architectural optimization target, it is also an abstraction of thermal constraints. 
Other factors impacting thermals include frame geometry and mating interfaces, component and frame materials, heat spreaders, and the physical placement of components. 
Notably, device thermals must be designed within thermal limits—especially those related to skin touch temperature—to be comfortable and safe for extended wear. 
While IEC 62368-1~\cite{iec62368-1} dictates a safety limit on touch temperature of 43–48°C, this does not meet the \textit{comfort} requirements for glasses~\cite{alaghi_cvpr_2022}. 

Because glasses either touch or reside near sensitive areas of the face, like the eyebrow, nose bridge, and temple, lower temperature limits beyond regulatory standards are needed in practice.
While the sustained power budget depends on the overall industrial and mechanical design, as well as environmental factors such as ambient temperature and solar heating, generally a touch temperature limit of \textasciitilde 40°C for comfort implies an upper limit for sustained power dissipation of approximately 1–2 watts. Thus, we expect that any work addressing thermals—whether through cooling solutions, temperature impacts of power reduction, or other thermal management strategies—will be valuable for future wearable devices.

\subsection{Evolving Workloads, Architectures, and Bottlenecks}

The system architecture, workloads, and power results presented in this paper represent a snapshot in time of a wearable contextual AI system. As these systems mature, the workloads for egocentric primitives and their resource demands will continue to evolve. Next-generation workloads may introduce new sensing modalities and compute requirements, while existing implementations may become more efficient through optimizations or less efficient due to novel or increased accuracy targets. These changes will continue to impact and shift the trade-offs between on-device compute and communication power, as well as system resource usage and contention.

We also expect industry trends and advancements to continue shifting power bottlenecks around over time. Advances in semiconductor manufacturing will continue to improve and redistribute these bottlenecks. At the same time, new workloads will require optimizing existing architectural resources, adding new ones, or sharing existing ones. Tracking how these developments affect overall system power and bottlenecks will be important for guiding system architecture design decisions going forward to avoid Amdahl's limitations applied to power.

This underscores the need for continuous and comprehensive system modeling to evaluate whether optimizations and future design investments reduce system power in full system context. Without a full system view to measure and monitor shifting system bottlenecks, research investments may end up optimizing components with minimal impact on the overall power consumption. Therefore, continuous monitoring of how workload and architectural trends will be essential to guide the development of future wearable system architectures.

\section{Conclusion}

\label{sec:conclusion}

Enabling the next generation of always-on contextual AI systems will require increasingly challenging improvements to system power. Our work defines a set of egocentric primitives to guide the design of current-generation contextual AI systems and insights for future systems. We build and demonstrate the value of a full-system architecture model to analyze design decisions and trends in full system context. In particular, we show the importance of optimizing the compute-communication power trade-off by balancing whether egocentric primitives run on or off device. On the road ahead, we expect egocentric primitives and device architectures to evolve over time, causing bottlenecks to shift throughout the system. Thus, it will be important to continuously evaluate design decisions in the full system context to ensure that Amdahl's Law for power does not limit the potential and progress for future contextual AI wearable systems.

\bibliographystyle{IEEEtranS}
\bibliography{references}

\end{document}